\documentclass[a4paper,12pt,reqno]{amsart}

\usepackage{amssymb}
\usepackage{amsfonts}
\usepackage{latexsym}

\usepackage[english]{babel}

\makeatletter
\newtheorem{theorem}{Theorem}

\newtheorem{e-proposition}[theorem]{Proposition}

\newtheorem{e-definition}[theorem]{Definition\rm}
\newtheorem{remark}{\it Remark\/}

\newcommand{\SLE}{{\mathrm{SLE}}}

\newcommand{\eg}{\emph{e.g.}}
\newcommand{\ie}{\emph{i.e.}}

\newcommand{\decr}{\mathop{\makebox[0pt][l]{\kern0.5em$\downarrow$}\bigcap}}

\newcommand{\incr}{\mathop{\makebox[0pt][l]{\kern0.5em$\uparrow$}\bigcup}}

\newcommand{\eps}{\varepsilon}

\begin{document}

\title{Convergence of Ising interfaces to~Schramm's~SLE~curves}

\author[D.~Chelkak]{Dmitry Chelkak$^\mathrm{a,b}$}
\author[H.~Duminil-Copin]{Hugo Duminil-Copin$^\mathrm{c}$}
\author[C.~Hongler]{Cl\'ement Hongler$^\mathrm{d}$}
\author[A.~Kemppainen]{Antti Kemppainen$^\mathrm{e}$}
\author[S.~Smirnov]{Stanislav Smirnov$^\mathrm{a,c}$}


\thanks{\textsc{${}^\mathrm{A}$ Chebyshev Laboratory, Department of Mathematics and Mechanics,
St.~Petersburg State University. 14th Line, 29b, 199178 St.~Petersburg, Russia}}
\thanks{\textsc{${}^\mathrm{B}$ St.~Petersburg Department of Steklov Mathematical Institute (PDMI RAS). Fontanka~27, 191023 St.~Petersburg, Russia}}
\thanks{\textsc{${}^\mathrm{C}$ Section de Math\'ematiques, Universit\'e de Gen\`eve. 2-4 rue du Li\`evre, Case postale~64, 1211 Gen\`eve 4, Suisse}}
\thanks{\textsc{${}^\mathrm{D}$ Department of Mathematics, Columbia University. 2990 Broadway, New York, NY 10027, USA}}
\thanks{\textsc{${}^\mathrm{E}$ Department of Mathematics and Statistics, University of Helsinki. P.O.~Box~68, FIN-00014, Finland}}

\thanks{{\it E-mail addresses:} \texttt{dchelkak@pdmi.ras.ru} (Dmitry Chelkak), \texttt{hugo.duminil@unige.ch} (Hugo Duminil-Copin), \texttt{hongler@math.columbia.edu} (Cl\'ement Hongler), \texttt{antti.h.kemppainen@helsinki.fi} (Antti Kemppainen), \texttt{stanislav.smirnov@unige.ch} (Stanislav Smirnov)}

\begin{abstract}
We show how to combine our earlier results to deduce strong convergence
of the interfaces in the planar critical Ising model and its random-cluster representation
to Schramm's SLE curves with parameter $\kappa=3$ and $\kappa=16/3$ respectively.
\end{abstract}

\maketitle

\section{Introduction and statement of the main theorems}
\label{}

In \cite{Sch00} Oded Schramm introduced SLE -- a family of random
fractal curves parameterized by $\kappa>0$, which are obtained by running Loewner evolution with a speed $\kappa$ Brownian motion as the driving term.
Schramm showed that those are the only possible conformally invariant scaling limits of interfaces in 2D~critical lattice models,
and the convergence to SLE was indeed proved in a number of cases, see~\cite{Law05book,Smi06}.

The 2D Ising model is one of the most studied models of an order-disordered phase transition.
Existence of a conformally invariant scaling limit at criticality 
in the sense of correlation functions was postulated in the seminal physics paper \cite{BPZ84a} and used to deduce unrigorously many of its properties since.
Recently, one of us \cite{Smi06} has constructed discrete holomorphic observables in the critical Ising model on bounded discrete domains (and its random cluster representation), which have been shown to have conformally invariant scaling limits in \cite{Smi10,CS12}.
This paved a way to an ongoing project of rigorously establishing conformally covariant scaling limits
for all critical Ising correlation functions, \cite{HS10,Hon10,CI13,CHI12}. 
On the other hand, it had a corollary that
interfaces in the spin Ising model and its FK (random cluster) counterpart
converge to SLE(3) and SLE(16/3) in the sense of the driving terms. Namely,
discrete interfaces are given by Borel probability measures on the space of driving terms with the uniform norm. As one passes to the scaling limit,
those measures converge weakly to the Brownian motions of the appropriate speeds.

To study geometric features of the interfaces, it is important to strengthen the topology of convergence to the uniform metric on the space of curves themselves. A possible way to arrive to this convergence was suggested in \cite{KS12},
with a corollary that FK interfaces converge to SLE(16/3) curves.
In this paper, we  provide a self-contained framework to establish this stronger convergence
both in the spin and the FK case, by combining the setup from \cite{KS12} with the crossing estimates from
\cite{CS12,CDCH13} and the results on the observables convergence from \cite{Smi10,CS12}.


\smallbreak
\noindent
{\em Definition of the model.} We discuss spin Ising and FK~Ising models on the square lattice ${\mathbb Z}^2$ (see \cite{DCS12} and references therein for background). The results can be generalized to isoradial planar lattices as explained in~\cite{CS12}.
For a finite subgraph $G=(V,E)\subset {\mathbb Z}^2$, the spin Ising model on $G$ at inverse temperature $\beta$ is a random assignment of spins $\sigma_x\in\{-1,+1\}$ such that the configuration $\sigma=(\sigma_x)_{x\in V}$ has probability proportional to $\exp(\beta\sum_{(xy)\in E}\sigma_x\sigma_y)$. The FK~Ising model is its random-cluster counterpart obtained via the Edwards-Sokal coupling, see \eg~\cite{DCS12}. More precisely, it is a dependent bond percolation model on $G$: the probability of a configuration $\omega\subset E$ is proportional to \mbox{$[p/(1-p)]^{{\mathrm{o}(\omega)}}2^{\,{\rm c}(\omega)}$}, where $\mathrm{o}(\omega)$ and $\mathrm{c}(\omega)$ are respectively the number of edges and connected components (\emph{clusters}) in $\omega$, and where \mbox{$p=1-e^{-2\beta}$}. We consider the models at the critical point $\beta_\mathrm{crit}=\frac{1}{2}\log(1\!+\!\sqrt 2)$ and $p_\mathrm{crit}=\sqrt {2}/(1\!+\!\sqrt{2})$.

\smallbreak
\noindent
{\em Interfaces for Dobrushin boundary conditions.} Let $\Omega$ be a bounded simply connected domain and \mbox{$a,b\in\partial\Omega$} be two distinct boundary points (more accurately, two degenerate prime ends, see \cite[\S 2.4,2.5]{Pom92book}) of~$\Omega$. We aim to approximate $\Omega$ (in any reasonable sense) by subgraphs of the square grids $\delta\mathbb{Z}^2$ successively refined as $\delta\to 0$.
Let $\Omega^\delta\subset \delta{\mathbb Z}^2$ be a simply connected (meaning connected and with connected complement) approximation, and $a^\delta$, $b^\delta$ be two vertices near $a$ and $b$ on the boundary $\partial \Omega^\delta$. When going in counterclockwise order, $a^\delta$ and $b^\delta$ define two arcs of $\partial\Omega^\delta$ denoted by $(a^\delta b^\delta)$ and $(b^\delta a^\delta)$.

For the spin Ising model, the boundary conditions  ``$-1$'' on $(a^\delta b^\delta)$ and ``$+1$'' on $(b^\delta a^\delta)$ are called \emph{Dobrushin boundary conditions} in $(\Omega^\delta;a^\delta,b^\delta)$. These boundary conditions generate a \emph{spin interface} $\gamma^\delta$ -- simple curve running from $a^\delta$ to $b^\delta$ that has spins ``$+1$'' on its left side and spins ``$-1$'' on its right.
For technical reasons (\eg~to avoid self-touchings), we prefer to draw $\gamma^\delta$ on the auxiliary square-octagon lattice, with octagons corresponding to the vertices of $\Omega^\delta$. We assume $\gamma^\delta$ to be the rightmost (or the leftmost) interface, but it could also turn arbitrarily in ambiguous situations with four alternating spins around a face. As we obtain the same limit for all choices of $\gamma^\delta$, the possible differences are only microscopic.

For the FK~Ising model, we consider free boundary conditions on $(a^\delta b^\delta)$ and wired ones on $(b^\delta a^\delta)$, and call them Dobrushin boundary conditions in $(\Omega^\delta;a^\delta,b^\delta)$. In this case, configurations are best seen together with their dual counterparts defined on the dual graph $G^*$. 
In the dual model
(which is again the critical FK Ising model)
the boundary conditions become
dual-wired on $({a}^\delta{b}^\delta)$ and dual-free on $({b}^\delta{a}^\delta)$. Let \mbox{$\gamma^\delta$} be the unique interface (again, drawn on the auxiliary square-octagon lattice, see~\cite[Section~4.1]{KS12}) that separates the FK cluster on ${G}$ connected to $({b}^\delta{a}^\delta)$ and the FK cluster on ${G}^*$ connected to $(a^\delta b^\delta)$.


\smallbreak \noindent {\em Statement of the theorems.}
We equip the space of continuous oriented curves by the following metric:
\begin{equation} \label{CurveDist}
d(\gamma_1,\gamma_2)~=~\inf\nolimits_{\phi_1,\phi_2} ||\gamma_1\circ\phi_1-\gamma_2\circ\phi_2||_\infty,
\end{equation}
where the infimum is taken over all orientation-preserving reparameterizations of $\gamma_1$,$\gamma_2$.

\smallbreak

\begin{theorem}[Convergence of spin~Ising interfaces]\label{convergence Ising interface}
Let $\Omega$ be a bounded simply connected domain with two distinct boundary points (degenerate prime ends) $a$,$b$. Consider the interface $\gamma^\delta$ in the critical spin Ising model with Dobrushin boundary conditions on $(\Omega^\delta;a^\delta,b^\delta)$. The law of $\gamma^\delta$ converges weakly, as~$\delta\rightarrow 0$, to the chordal Schramm-Loewner Evolution $\mathrm{SLE}(\kappa)$ running from $a$ to $b$ in $\Omega$ with $\kappa=3$.
\end{theorem}

\smallbreak

\begin{theorem}[Convergence of FK~Ising interfaces]\label{convergence FK interface}
Let $\Omega$ be a bounded simply connected domain with two distinct boundary points (degenerate prime ends) $a$,$b$. Consider the interface $\gamma^\delta$ in the critical FK~Ising model with Dobrushin boundary conditions on $(\Omega^\delta;a^\delta,b^\delta)$. The law of $\gamma^\delta$ converges weakly, as~$\delta\rightarrow 0$, to the chordal Schramm-Loewner Evolution $\mathrm{SLE}(\kappa)$ running from $a$ to $b$ in $\Omega$ with $\kappa=16/3$.
\end{theorem}

\smallbreak


\smallbreak
\noindent
{\em Chordal Loewner evolution.} Below we briefly explain the construction of Schramm's $\SLE$ curves and introduce the notation which is used in the next sections (see \cite{Law05book} for further background).
Let $\gamma_\mathbb{D}$ be some continuous non-self-crossing (though maybe self-touching) curve running in the closed unit disc $\overline{\mathbb{D}}$ and parameterized by $s\in [0,1]$ such that $\gamma_\mathbb{D}(0)=-1$ and $\gamma_\mathbb{D}(1)=+1$. Let
\mbox{$\Phi:z\mapsto i\cdot{(1\!+\!z)}/{(1\!-\!z)}:\mathbb{D}\to\mathbb{H}$} be the fixed conformal map from $\mathbb{D}$ onto the upper half-plane $\mathbb{H}$ and $\gamma_\mathbb{H}=\Phi(\gamma_\mathbb{D})$, thus
$\gamma_\mathbb{H}$ starts at $0$ and goes to $\infty$. Denote by ${K}_s$ the hull of $\gamma_\mathbb{H}[0,s]$, \ie~the complement of the connected component of $\overline{\mathbb{H}}\setminus \gamma_\mathbb{H}[0,s]$ containing~$\infty$,
and let $t(s)=\mathrm{hcap}(K_s)$ be the half-plane capacity of $K_s$. It is easy to see that $t(s)$ is nondecreasing but there could be situations when it remains constant (and, moreover, the hulls $K_s$ remain the same) for a nonzero time, even if $\gamma_\mathbb{D}$ is obtained as a limit of simple curves. \emph{E.g.}, {\bf (a)} it might happen that, for some $s\in (0,1)$, the tip $\gamma_\mathbb{H}(s)$ of the growing curve is not visible from $\infty$ (meaning that $\gamma_\mathbb{H}$ explores some inner component of $\mathbb{H}\setminus \gamma_\mathbb{H}([0,s])$) or {\bf (b)} $\gamma_\mathbb{H}(s+\cdot)$ might travel along the boundary of $K_s$ for a nonzero time, not changing the hull. Also, it might happen that {\bf (c)} $\gamma_\mathbb{D}$ reaches $+1$ for the first time before $s=1$ or {\bf (d)} $t(s)$ remains bounded as $s\to 1$ (if $\gamma_\mathbb{H}$ goes to $\infty$ very close to $\mathbb{R}$). We say that

\smallbreak

\noindent $\gamma_\mathbb{D}$ \emph{can be fully described by the Loewner evolution} if none of (a)-(d) happens and $t(s)$ is strictly increasing.

\smallbreak

\noindent In this situation, let $g_t:\mathbb{H}_t=\mathbb{H}\setminus {K}_{s(t)}\to \mathbb{H}$ be the conformal map such that $g_t(w)=w+2t\cdot w^{-1}+O(w^{-2})$ as $w\to\infty$. The Loewner equation is
\begin{equation}
\label{LoewnerEq}
\frac{dg_t(w)}{dt}=\frac{2}{g_t(w)-W_t},\quad w\in\mathbb{H}_t, 
\end{equation}
where $W_t$ is a continuous function which is usually called a driving term. Schramm-Loewner Evolutions -- $\SLE$s for short -- are the random curves constructed in the upper half-plane by solving the Loewner equation with $W_t=\sqrt{\kappa}B_t$, $\kappa>0$, and then defined in all other simply connected domains $(\Omega;a,b)$ with two marked boundary points (\eg~in the unit disc $(\mathbb{D};-1,+1)$) via conformal maps.

\section{Tightness and crossing bounds}
\label{sec:tightness}
The proof of the main theorems starts with the extraction of subsequences from the laws of the discrete interfaces in the topology associated to the metric (\ref{CurveDist}). By recent results of two of us~\cite{KS12} (which strengthen classical results of~\cite{AB99}) the tightness of the family $\{\gamma^\delta\}$ follows once we can guarantee a crossing estimate called Condition~G, see below. This condition also guarantees that subsequential limits of $\gamma^\delta$ can be fully described by the Loewner evolution with driving terms having finite exponential moments.

Let $\Omega^\delta_\mathbb{C}\subset\mathbb{C}$ be the polygonal domain (union of tiles) corresponding to $\Omega^\delta\subset\delta\mathbb{Z}^2$ and $\phi^\delta:(\Omega^\delta_\mathbb{C}; a^\delta, b^\delta)\to (\mathbb{D}; -1,+1)$ be some conformal maps. Note that until Section~\ref{sec:convergence} we do not need to normalize $\phi^\delta$ in any specific way.

\smallbreak

\begin{theorem}[\cite{KS12}] \label{KemSmiThm}
Let $\Omega$ be a bounded simply connected domain with two distinct degenerate prime ends $a$ and~$b$. If the family of probability measures $\{\gamma^\delta\}$ satisfies the Condition G given below, then both $\{\gamma^\delta\}$ and $\{\gamma^\delta_\mathbb{D}\}$ are tight in the topology associated with the curve distance~(\ref{CurveDist}). Moreover, if $\gamma^\delta_\mathbb{D}$ is converging weakly to some random curve $\gamma_\mathbb{D}$, then the following statements hold:
\begin{enumerate}
\item a.s., the curve $\gamma_\mathbb{D}$ can be fully described by the Loewner evolution and the corresponding maps $g_t$ satisfy the equation (\ref{LoewnerEq}) with a driving process $W_t$ which is $\alpha$-H\"older continuous for any $\alpha < \frac{1}{2}$;
\item the driving processes $W_t^\delta$ corresponding to $\gamma^\delta_\mathbb{D}$ converge in law to $W_t$ with respect to the uniform norm on finite intervals; moreover, $\sup_{\delta>0}\mathbb{E}[\exp(\eps |W_t^\delta|/\sqrt{t}\,)] < \infty$ for some $\epsilon>0$ and all $t$.
\end{enumerate}
\end{theorem}

\smallbreak

\begin{remark}
The theorem combines  several results from \cite{KS12}. Note that, if the prime ends $a,b$ are degenerate, the convergence of $\gamma^\delta$ outside of their neighborhoods implies the convergence of the whole curves. 
\end{remark}

\smallbreak
\noindent
{\em Crossing bounds.} We say that a curve $\gamma^\delta$ makes a \emph{crossing} of an annulus $A(z_0,r,R)=B(z_0,R)\setminus \overline{B(z_0,r)}$, if it intersects both its inner and outer boundaries $\partial B(z_0,r)$ and $\partial B(z_0,R)$. We say that the crossing is \emph{unforced} if it can be avoided by deforming the curve inside of $\Omega^\delta_\mathbb{C}$; in other words, if it occurs along a subarc of $\gamma^\delta$ contained in a connected component of $A(z_0,r,R)\cap\Omega^\delta_\mathbb{C}$ that does not disconnect $a^\delta$ and $b^\delta$.

\smallbreak

\noindent {\bf Condition~G} \emph{The curves $\gamma^\delta$ are said to satisfy a \emph{geometric bound on unforced crossings}, if
there exists $C>1$ such that, for any $\delta>0$ and any annulus $A(z_0,r,R)$ with $R/r>C$ such that $\partial B(z_0,r)\cap\partial \Omega^\delta_\mathbb{C}\ne \emptyset$,}

\smallbreak

\begin{center}
$\mathbb{P}[\gamma^\delta~\mathit{makes~an~unforced~crossing~of}~A(z_0,r,R)]< {\textstyle \frac{1}{2}}$.
\end{center}

\smallbreak

\begin{remark}
Actually, the results of \cite{KS12} are based on the stronger Condition~G2: the similar crossing bound should hold at any stopping time $\tau$.
As our interfaces $\gamma^\delta$ satisfy the domain Markov property ($\gamma^\delta$ after time $\tau$ has the same distribution as the interfaces in the slit domains $\Omega^\delta\setminus\gamma^\delta[0,\tau]$), it is sufficient to check ``time zero'' Condition~G \emph{for all domains $\Omega^\delta$ simultaneously}, see \cite[Section~2]{KS12} for further discussion.
\end{remark}

\smallbreak

Condition G deals with crossings of the simplest possible geometric shapes but it is not clear a priori if it is stable under conformal maps. One of the ways to prove this fact is to use a larger class of shapes. Namely, instead of annuli one can consider all \emph{conformal rectangles} $Q$, \ie~conformal images of rectangles $\{z:\mathrm{Re}\,z\in(0,\ell), \mathrm{Im}\,z\in (0,1)\}$. For a given $Q$, we call ``marked sides'' the images of the segments $[0,i]$ and $[\ell,\ell+i]$ and ``unmarked'' the other two sides, and call the (uniquely defined) quantity $\ell=\ell(Q)$ the \emph{extremal length} of $Q$. We say that $\gamma^\delta$ makes a crossing of $Q$ if $\gamma^\delta$ intersects both of its marked arcs.

\smallbreak

\noindent {\bf Condition~C} \emph{The curves $\gamma^\delta$ are said to satisfy a \emph{conformal bound on unforced crossings} if there exist $L,\eta>0$ such that, for any $\delta>0$ and any conformal rectangle $Q\subset\Omega^\delta_\mathbb{C}$ that does not disconnect $a^\delta\!$ and $b^\delta\!$,}

\smallbreak

\begin{center}
\noindent \emph{if $\ell(Q)>L$ and the unmarked sides of $Q$ lie on $\partial \Omega^\delta_\mathbb{C}$, then $\mathbb{P}\big[\gamma^\delta~\mathit{makes~a~crossing~of}~Q\big]<1-\eta$.}
\end{center}

\smallbreak

\begin{remark} It is shown in~\cite{KS12} that Conditions~G and~C are equivalent, in particular Condition~G is conformally invariant.  Thus, if these conditions hold for the curves $\gamma^\delta$, then they hold for $\gamma^\delta_\mathbb{D}$ too.
\end{remark}

\smallbreak

There are two approaches to check that interfaces $\gamma^\delta$ fit within the setup described above. The first is straightforward: to derive the needed uniform estimate for all shapes (parts of annuli or conformal rectangles), including those with irregular boundaries. Recently, three of us have proved such an estimate, following the ideas from~\cite{DCHN11} and relying on the new discrete complex analysis techniques developed in~\cite{Che12}.

\smallbreak

\begin{theorem}[\cite{CDCH13}] \label{ThmCrossing} For any $L>1$ there exist $\eta>0$ such that for any discrete domain $(\Omega^\delta;a^\delta,b^\delta,c^\delta,d^\delta)$ with four marked boundary points and
\mbox{$L^{-1}<\ell_{\mathrm{d}}(\Omega^\delta;(a^\delta b^\delta),(c^\delta d^\delta))<L$}, one has
\smallbreak
\begin{center}
$\eta<\mathbb{P}[\mathit{there~is~an~FK~cluster~connecting}~(a^\delta b^\delta)~\mathit{and}~(c^\delta d^\delta)~\mathit{inside~of}~\Omega^\delta]<1-\eta$
\end{center}
\smallbreak
\noindent uniformly over all possible boundary conditions on $\partial\Omega^\delta$, where $\ell_{\mathrm{d}}$ denotes the discrete extremal length.
\end{theorem}

\smallbreak

Note that, in order to verify Condition~C for FK Ising interfaces in polygonal domains $\Omega^\delta_\mathbb{C}$,
one does not need to consider very small conformal rectangles near $\partial\Omega^\delta_\mathbb{C}$ as \emph{$\gamma^\delta$ never visits $\frac{1}{4}\delta$ neighborhood of~$\partial\Omega^\delta_\mathbb{C}$} due to our choice of the square-octagon lattice as the graphical representation of the model, see~\cite[Section~4.1.4]{KS12}. 
Thus, the result follows from Theorem~\ref{ThmCrossing} since $\ell_{\mathrm{d}}(Q)$ and $\ell(Q)$ are uniformly comparable.

The second approach is to use the monotonicity of crossing probabilities for specific boundary conditions. In~\cite[Section~4.1.6]{KS12}, two of us have shown that it is enough to consider only two particular types of regular annulus-like shapes with alternating wired/free/wired/free boundary conditions. In this case, the needed estimate can be easily extracted from \cite[Theorem~1.3]{CS12}. It is worthwhile noting that two approaches described above have different advantages: the second does not require hard technicalities while the first can be applied in more general situations, \eg~to the analysis of branching interface trees.

\smallbreak

\begin{remark}
As the crossing bound for the FK model with alternating boundary conditions is proved, the upper bound for the probability of a ``$+1$'' crossing in the spin model with ``$+1$/free/$-1$/free/$+1$/free/$-1$/free'' boundary conditions (and so, by monotonicity, with ``$+1$/$-1$/$+1$/$-1$'' ones) can be derived using the Edwards-Sokal coupling, see~\cite{CDCH13}.
\end{remark}

\section{Identification of the limit via convergence of martingale observables}
\label{sec:convergence}

We identify the scaling limits of $\gamma^\delta$ along subsequences following the generalization of the approach from~\cite{LSW04}, outlined in~\cite{Smi06}. It
requires the identification of the scaling limit of a non-trivial martingale observable, known so far for but a few lattice models.
We sketch the proof of Theorem~\ref{convergence Ising interface} starting with \cite[Theorem~1.2]{CS12}, the similar derivation of Theorem~\ref{convergence FK interface} from \cite[Theorem~2.2]{Smi10} can be found in~\cite[Section~6.3]{DCS12}.

From now onwards we fix the conformal maps \mbox{$\phi:(\Omega;a,b)\to(\mathbb{D};-1,+1)$} and \linebreak \mbox{$\phi^\delta:(\Omega^\delta_\mathbb{C};a^\delta,b^\delta)\to(\mathbb{D};-1,+1)$} so that $\phi^\delta(z)\to\phi(z)$ as $\delta\to 0$ uniformly on compact subsets of~$\Omega$. Let $w(z)=\Phi(\phi(z))$ and $w^\delta(z)=\Phi(\phi^\delta(z))$.

\smallbreak

\noindent \emph{Proof of Theorem~\ref{convergence Ising interface}} Let $F^\delta_n$ be the discrete fermionic observable in the domain $(\Omega^\delta_n;\gamma^\delta_n,b^\delta)$, appropriately normalized at $b^\delta$, where $\Omega^\delta_n$ denotes the connected component of the slit domain $\Omega^\delta_\mathbb{C}\setminus (\gamma^\delta_0\gamma^\delta_1...\gamma^\delta_n)$ containing $b^\delta$, see \cite[Section~2.2.1]{CS12}. Let $g^\delta_t:\mathbb{H}_t^\delta=\Phi(\phi^\delta(\Omega^\delta_n))\to\mathbb{H}$ be the corresponding Loewner evolutions with driving terms $W^\delta_t$, reparameterized by the capacity. Theorems~1.2 and~5.6 of \cite{CS12} state that

\smallbreak

\begin{center}
\emph {$|F^\delta_n(z)- M_t^\delta(z)|\to 0$ as $\delta\to 0$ uniformly over all possible domains $\Omega^\delta_n$ and all $z$ in\\ the bulk of $\Omega^\delta_n$, where $M_t^\delta(z)= (\partial_z [-G_t^\delta(w^\delta(z))^{-1}]\,)^{1/2}$  and $G_t^\delta(w)=g_t^\delta(w)-W_t^\delta$.}
\end{center}

\smallbreak

\noindent Recall that, for a given $\delta>0$ and $z^\delta$ in $\Omega^\delta$, the value $F^\delta_n(z^\delta)$ is a martingale with respect to the filtration $(\mathcal{F}^\delta_n)_{n\ge 0}$, where $\mathcal{F}^\delta_n$ is generated by the first $n$ steps of~$\gamma^\delta$. It is easy to see that, for all driving terms, $\mathrm{Im}\, G_t^\delta(w)\ge 2\sqrt{t}$, as long as $\mathrm{Im}\, w\ge 3\sqrt{t}$. In particular, $M_t^\delta(z)$ are uniformly bounded and equicontinuous on compact subsets of $\Omega$, if $\delta$ is small enough and $t\le \frac{1}{9}(\mathrm{Im}\,w(z))^2$.

Theorem~\ref{KemSmiThm} gives us both the convergence of curves $\gamma^\delta$ and their driving processes $W_t^\delta$, at least along subsequences. Moreover, a.s., the limit can be fully described by the Loewner evolution with a continuous driving process $W_t$. Using
the convergence $w^\delta\to w$, the equicontinuity (in~$t$) of $g_t^\delta$
and the convergence of $G_t^\delta$ to $G_t$ in the bulk of $\mathbb{H}_t$ 
(which follows from the convergences of $W^\delta_t$ to $W_t$), we conclude that

\smallbreak

\begin{center}
\emph{for any $z\in\Omega$, the process $M_t(z)=(\partial_z [-G_t(w(z))^{-1}])^{1/2}$, $t\le T(z)$,\\ where $G_t(w)=g_t(w)-W_t$ and $T(z)=\frac{1}{9}(\mathrm{Im}\, w(z))^2$,\\ is a martingale with respect to the filtration $(\mathcal{F}_t)_{t\ge 0}$ generated by $W_t$.}
\end{center}

\smallbreak

Recall that $W_t$ is continuous a.s., and let $\tau=\tau(z)$ be the first time such that $\mathrm{Im}\, w(z)= 3(\sqrt{\tau}+|W_{\tau}|)$. Starting with the expansion $G_t(w)=w-W_t+2t\cdot w^{-1}+O(w^{-2})$ as $w\to\infty$, one directly gets
\begin{equation}\label{AsymptExpansion}
M_{t\wedge\tau}(z)=(w'_z)^{1/2}w^{-1}\cdot[1+W_{t\wedge\tau}\cdot w^{-1}+(W_{t\wedge\tau}^2-3(t\wedge\tau))\cdot w^{-2}+O(w^{-3})],\quad w=w(z),
\end{equation}
where the $O$-bounds are uniform with respect to both $t$ and $z$.
Since (\ref{AsymptExpansion}) is a martingale for \emph{any} given $z\in\Omega$ and $W_t$ has a finite exponential moment, we can exchange the asymptotic expansion with the conditional expectation and conclude that both coefficients $W_t$ and $W_t^2-3t$ are martingales.
As $W_t$ is almost surely continuous, L\'evy's theorem 
implies that $W_t=\sqrt{3}B_t$, where $B_t$ is a standard Brownian motion, for any subsequential limit of the curves $\gamma^\delta$. \hfill $\qed$




\section*{Acknowledgements}
\noindent This research was supported by the Swiss NSF and ERC AG CONFRA. Dmitry Chelkak and Stanislav Smirnov were partly supported by the Chebyshev Laboratory at Saint Petersburg State University under the Russian Federation Government grant 11.G34.31.0026 and JSC~``Gazprom Neft''. Cl\'ement Hongler was partly supported by the National Science Foundation under grant DMS-1106588 and the Minerva Foundation. Antti Kemppainen was supported by the Academy of Finland.\\



\end{document}